\documentclass[3p]{elsarticle}

\usepackage{amsmath}
\usepackage{amsthm}
\usepackage{amssymb}
\usepackage{graphicx}
\usepackage{esint}
\usepackage{color}
\PassOptionsToPackage{hyphens}{url}
\usepackage[pdfencoding=auto,psdextra,colorlinks=true,pdfauthor=author]{hyperref}
\usepackage{bookmark}

\biboptions{numbers,sort&compress}

\journal{BioSystems}

\begin{document}

\begin{frontmatter}

\title{Quantum propensities in the brain cortex and free will}

\author[address1]{Danko D. Georgiev\corref{mycorrespondingauthor}}
\ead{danko.georgiev@mail.bg}
\cortext[mycorrespondingauthor]{Corresponding author}

\address[address1]{Institute for Advanced Study, 30 Vasilaki Papadopulu Str., Varna 9010, Bulgaria}

\begin{abstract}
Capacity of conscious agents to perform genuine choices among future alternatives is a prerequisite for moral responsibility. Determinism that pervades classical physics, however, forbids free will, undermines the foundations of ethics, and precludes meaningful quantification of personal biases. To resolve that impasse, we utilize the characteristic indeterminism of quantum physics and derive a quantitative measure for the amount of free will manifested by the brain cortical network. The interaction between the central nervous system and the surrounding environment is shown to perform a quantum measurement upon the neural constituents, which actualize a single measurement outcome selected from the resulting quantum probability distribution. Inherent biases in the quantum propensities for alternative physical outcomes provide varying amounts of free will, which can be quantified with the expected information gain from learning the actual course of action chosen by the nervous system. For example, neuronal electric spikes evoke deterministic synaptic vesicle release in the synapses of sensory or somatomotor pathways, with no free will manifested. In cortical synapses, however, vesicle release is triggered indeterministically with probability of 0.35 per spike. This grants the brain cortex, with its over 100 trillion synapses, an amount of free will exceeding 96 terabytes per second. Although reliable deterministic transmission of sensory or somatomotor information ensures robust adaptation of animals to their physical environment, unpredictability of behavioral responses initiated by decisions made by the brain cortex is evolutionary advantageous for avoiding predators. Thus, free will may have a survival value and could be optimized through natural selection.
\end{abstract}


\date{July 1, 2021}

\end{frontmatter}

%\linenumbers

%\tableofcontents

\section*{Highlights}

\begin{itemize}

\item Free will is the capacity of conscious agents to choose a future course of action among several available physical alternatives.
	
\item Expected information gain from learning the choice of a conscious agent provides a quantitative measure of free will.
	
\item Quantum indeterminism supports varying amounts of free will exercised in different quantum measurement contexts.
	
\item Probabilistic release of synaptic vesicles from cortical synapses grants on average 0.934 bits of free will per synapse per spike.
	
\item Unpredictability of animal behavior provides a survival advantage and allows for evolutionary optimization of manifested free will.
	
\end{itemize}

\section{Introduction}

We are conscious beings who feel in control of their future actions
\citep{Georgiev2017}. Good choices make us happy and elevate well-being,
whereas bad choices make us regret about missed opportunities and
precipitate suffering. Importantly, the choices that we make on a
daily basis impact not only our lives, but also the lives of others
who depend on us. Consequently, the capacity of being able to choose
freely, comes with a moral responsibility and accountability for our
own actions \citep{Kane1996,Kane1999,Kane2005}. Thus, free will appears
to be a main prerequisite for personal fulfillment, through undertaking
steps toward achievement of individual life goals \citep{Sartre1965,Sartre2007}, and the construction
of a moral and just society, through legislation of civil law that
guarantees equal basic rights to all society members \citep{Rawls1958}.

Freedom is a basic human right and everyone instinctively strives
for a life without external coercion. We enjoy being free and the
existence of free will is persistently corroborated by our own introspective
testimony, e.g., whenever we decide to move our arms or legs \citep{Georgiev2017}. Thus,
it may come as a surprise that a neuroscientist could even consider
challenging the veracity of free will, which our conscious experiences
reveal to us. Moreover, regardless of how impeccable the logical reasoning
against free will may be, in ordinary circumstances such a directly
verifiable fact should be impervious to any argumentation that goes
against it. Instead, arriving at a contradiction should be considered
as a proof against the veracity of the set of initial assumptions
upon which the logical reasoning is based \citep{Georgiev2017}. Nonetheless,
it appears that what would be considered abnormal in ordinary circumstances
is accepted as quite normal in the philosophy of free will \citep{Lazerowitz1984}.
From the deterministic nature of physical laws in classical mechanics
\citep{Susskind2013}, it has been often concluded that free will is
impossible and we believe in an illusion \citep{Schopenhauer1841,Wegner2003,Nichols2011,Harris2012}.
Because the physical laws hold true at all times and do not evolve,
it appears that free will cannot evolve too, namely, either we have
free will or we do not, by virtue of physical laws. Indeed, if physical
laws always forbid the existence of free will, there should be no
sense in which we acquire free will gradually through natural selection.
Furthermore, the concept of free will, defined as the capacity of
agents to choose a course of action among at least two alternative
future possibilities, appears to be a rough qualitative statement
that is unable to account for the possible presence of inherent biases
in favor of some of the available choices. Here, we will show that
free will can evolve in physical theories that allow a continuous
amount of free will to be exercised, that is from completely unbiased to completely
biased choosing. Then, we will demonstrate that quantum physics permits free will, whereas classical physics does not.

To address the problem of free will and its neurophysiological support in the brain, we first introduce a precise quantitative measure of free will based on the classical bits of Shannon information gained when an agent makes a choice at points of bifurcation in its physical dynamics (Section~\ref{sec:Shannon}).
Next, we scrutinize the intractability of free will within classical physics and pinpoint its origin in the deterministic Hamilton's equations (Section~\ref{sec:Classical}).
Then, we explain how modern quantum theory furnishes a physical measurement process in which conscious agents are able to exercise their free will (Section~\ref{sec:Quantum}),
and elaborate on the neurophysiological mechanisms that could evolve through natural selection to harness the possible biasing/unbiasing of inherent quantum propensities as provided by quantum indeterminism (Section~\ref{sec:Neurophysiology}).
Further, we explore the implications of the presented quantum approach for personal responsibility,
ethics, and moral values (Section~\ref{sec:Moral}).
Lastly, we conclude with a discussion (Section~\ref{sec:Discussion}) on synaptic learning through trial-and-error mechanism and
elucidate how positive or negative feedback could affect the amount of free will possessed by synapses in the brain cortex.

\section{Quantitative measure of free will}
\label{sec:Shannon}

\begin{figure}[t]
\begin{centering}
\includegraphics[width=162mm]{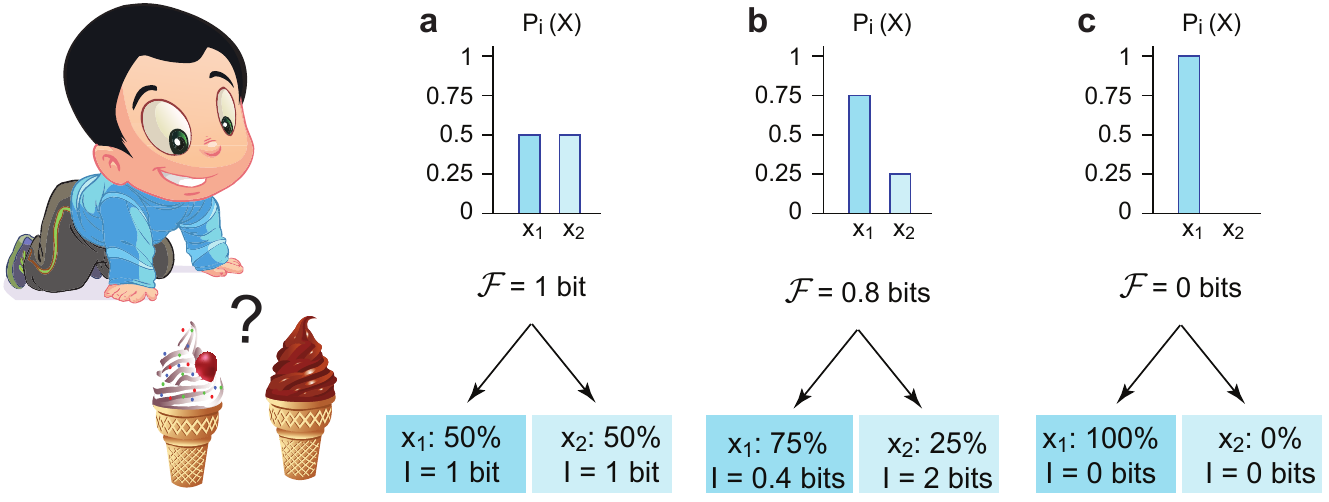}
\par\end{centering}
\caption{\label{fig:1}The amount of free will exercised by a conscious agent
could be quantified by the expected (average) information gain from
learning the actual choice performed by the agent. For illustration
is used the simplest case of a probability distribution~$P_{i}(X)$ with
only two outcomes $x_{1}$ and $x_{2}$, where the corresponding probabilities are interpreted as inherent propensities.
If the two initial probabilities are equiprobable, namely, both are 50\%, then each choice delivers
1\,bit of new information indicating that the act of choosing was
unbiased and completely free~(a). However, if one of the initial probabilities
is larger, 75\% vs 25\%, then the average choice delivers only 0.8\,bits
of new information indicating that the act of choosing was biased
and only partially free~(b). When one of the initial probabilities
is absolutely certain, 100\% vs 0\%, then the choice delivers no new
information indicating that the act of choosing was not free at all~(c).
Legend: $\mathcal{F}$~denotes the amount of free will, whereas $I$~denotes the information gain for each particular outcome.}
\end{figure}

Free will is characterized by the capacity of conscious agents to
make a genuine choice among several (at least two) alternative future courses of action.
The ability to do otherwise is an essential ingredient of the act
of choosing \citep{Kane1996,Kane1999,Kane2000b,Kane2005,Kane2009,Kane2014b}.
Furthermore, the choice needs to be exercised in the absence of external
coercion in order to be free. Our internal desires, however, may influence
and bias the probabilities with which different alternatives are actualized.
For example, suppose that you are given the choice of having an ice
cream with either vanilla or chocolate flavor (Figure~\ref{fig:1}). 
If the two choice outcomes are equiprobable due to equal desires, namely, both are 50\%, then the choice is unbiased and completely free. However, if one of the probabilities is larger,
say 75\% for vanilla and 25\% for chocolate, then the choice is biased
and only partially free. When one of the probabilities becomes absolutely
certain, say 100\% for vanilla and 0\% for chocolate, then the choice
is not free at all.

Before the actual choice of the agent is executed, each of the above
three cases has a different probability distribution $P(X)$ with
corresponding \emph{Shannon entropy} $H(X)$ measured in bits \citep{Shannon1948a,Shannon1948b}
\begin{equation}
H(X)=-\sum_{k}P(x_{k})\log_{2}P(x_{k}), \label{eq:Shannon}
\end{equation}
where $X$ is the discrete random variable with possible outcomes
$x_{1},x_{2},\ldots,x_{k}$ that occur with probabilities $P(x_{1}),P(x_{2}),\ldots,P(x_{k})$.
The function $f(P)=-P\log_{2}P$ is concave within the unit interval
$P\in[0,1]$, where by convention $0=0\log_{2}0$ \citep{Cover2006}.

The \emph{information content} (also called \emph{surprisal}) of an
individual outcome $x_{k}$ is defined as
\begin{equation}
I(x_{k})=-\log_{2}P(x_{k}) . \label{eq:surprisal}
\end{equation}
Therefore, the Shannon entropy of a distribution $P(X)$ is the average
(expected) \emph{information content} of the outcome of a random trial
\begin{equation}
H(X)=\sum_{k}P(x_{k})I(x_{k}) .
\end{equation}

A closely related concept, which considers the dynamic update from
an initial probability distribution $P_{i}(X)$ to a final probability
distribution $P_{f}(X)$, is the \emph{information gain} $D(P_{f}\Vert P_{i})$
(also called \emph{Kullback--Leibler divergence} or \emph{discrimination
distance}) given by \citep{Kullback1951,Kullback1987}
\begin{equation}
D(P_{f}\Vert P_{i})=\sum_{k}P_{f}(x_{k})\log_{2}\left[\frac{P_{f}(x_{k})}{P_{i}(x_{k})}\right] . \label{eq:kullback}
\end{equation}
In the case when the final distribution is peaked onto a single outcome
$x_{k}$, namely, $P_{f}(x_{k})=1$, the information gain reduces to
the information content (surprisal) \eqref{eq:surprisal} resulting
from occurrence of the individual outcome $x_{k}$ 
\begin{equation}
D(P_{f}(x_{k})=1\Vert P_{i})=-\log_{2}P_{i}(x_{k}) .
\end{equation}
Consequently, weighted averaging over all possible single-outcome information gains
with their corresponding probabilities of occurrence $P_{i}(x_{k})$
returns the Shannon entropy of the initial probability distribution
\begin{equation}
\sum_{k}P_{i}(x_{k})D(P_{f}(x_{k})=1\Vert P_{i})=H(P_{i}(X)) .
\end{equation}
Therefore, the Shannon entropy $H(X)$ of a distribution $P(X)$ is
also the average (expected) \emph{information gain} from learning
the outcome of a random trial.

Consider now a conscious agent who is given the choice of having an
ice cream, where he or she is able to select either vanilla or chocolate
flavor. The \emph{expected information gain} from learning the actual
choice performed by the agent could be identified with the \emph{amount
of free will} $\mathcal{F}$ exercised by the agent. Thus, free will
is quantifiable through the Shannon entropy $H_{i}$ of the initial
probability distribution $P_{i}(X)$ determined by the physical laws
that govern the time dynamics of the agent up to the point of bifurcation
of future trajectories
\begin{equation}
\mathcal{F}=-\sum_{k}P_{i}(x_{k})\log_{2}P_{i}(x_{k}) . \label{eq:fw}
\end{equation}
The amount of free will $\mathcal{F}$ is measured in bits and represents
the expected (average) surprisal resulting from the act of choosing
(Figure~\ref{fig:1}). The main difference between \eqref{eq:fw}
and \eqref{eq:Shannon} is that the Shannon entropy $H(X)$ could
also be applied to quantify ``fictitious'' probability distributions
representing personal ignorance about the actual physical state of
the world \citep{Adami2016,Gell-Mann1996}, whereas the amount of free
will $\mathcal{F}$ is defined only with respect to the fundamental
physical laws that govern the dynamics of the physical system.

In an ideal setting allowing for the same act of choosing to be repeated multiple times, $t_{1},t_{2}\ldots,t_{n}$,
the existing biases will be manifested in the limit of large number of repetitions~$n$, where the relative frequency~${n_k}/{n}$ of the outcome $x_k$ (that occurred $n_k$ number of times) approaches the initial probability~$P_{i}(x_k)$, namely
\begin{equation}
\lim _{n\to \infty} \frac{n_k}{n}=P_{i}(x_k) .
\end{equation}

The quantitative measure of free will $\mathcal{F}$ given by \eqref{eq:fw} is consistent with our intuitive expectation that the amount of free will should be independent on whether we make two choices simultaneously or not. Indeed, suppose that we have to make two consecutive unbiased choices for type of ice cream (vanilla or chocolate) and type of container (waffle cone or plastic cup). In such case, the resulting amount of free will exercised is 2 bits, which is the sum of two 1-bit choices. Alternatively, if we have to choose one of the four available combinations of ice cream and container at once, then we will exercise a single 2-bit choice. Thus, the overall amount of free will is the same as in the case with two consecutive choices.

\section{Intractability of free will in classical physics}
\label{sec:Classical}

Determinism is the main characteristic feature of classical physics
\citep{Georgiev2017}. A physical theory is deterministic, if from
the current state $S(t=0)$ of a closed physical system, the physical
laws permit mathematical prediction of any future state $S(t>0)$
with absolute certainty and arbitrarily high precision. In other words,
probabilities in deterministic theories do not have a fundamental
physical origin, but can only reflect subjective ignorance with respect
to the objective physical reality of the surrounding world.

\paragraph{Phase space description}
In classical mechanics, the physical state of a system is mathematically
represented by a point in phase space \citep{Susskind2013}. The phase
space is a multi-dimensional abstract space in which every degree
of freedom of the physical system is represented by an axis. For a
composite system of $n$ particles, the phase space contains $3n$
canonical position coordinates $q_{i}$, representing the $x$, $y$,
$z$ position components of each particle, and $3n$ canonical momentum
coordinates $p_{i}$, representing the $p_{x}$, $p_{y}$, $p_{z}$
momentum components of each particle. The state of the system $S(q_{i},p_{i},t)$
evolves in time according to the system of Hamilton's equations
\begin{equation}
\frac{dq_{i}}{dt}=\frac{\partial H(q_{i},p_{i},t)}{\partial p_{i}},\quad\frac{dp_{i}}{dt}=-\frac{\partial H(q_{i},p_{i},t)}{\partial q_{i}}, \label{eq:Hamilton}
\end{equation}
where $d$ is is total derivative operator, $\partial$ is partial
derivative operator, and the Hamiltonian $H(q_{i},p_{i},t)$ is a
distinguished physical observable corresponding to the total energy
of the system.

Solving Hamilton's equations for a classically admissible Hamiltonian
shows that the dynamic trajectories of classical physical systems
are continuous paths in phase space that do not contain any genuine
bifurcation points. This implies that free will is impossible in classical
physics. The main culprit for this state of affairs is the incompatibility
between free will and determinism. Indeed, if the state of the physical
system is a point in phase space, the Shannon entropy of the initial
probability distribution is always zero, $H(P_{i}(X))=0$, and according
to \eqref{eq:fw} there is no information gain and no free will, $\mathcal{F}=0$.

The only way that probabilities can occur in classical physics is
due to ignorance of the initial state $S(t=0)$ of the physical system.
Suppose that our measurement instruments operate with some finite
precision and we can locate with probability $P$ the initial state
of measured physical system to a certain volume in phase space consisting
of all positions between $q_{i}$ and $q_{i}+\Delta q_{i}$ and all
momenta between $p_{i}$ and $p_{i}+\Delta p_{i}$. The probability
density is $\rho=P/V$, where the volume in phase space is $V=\prod_{i}\Delta q_{i}\Delta p_{i}$.
The resulting dynamic evolution from Hamilton's equations \eqref{eq:Hamilton}
obeys Liouville's theorem according to which the phase-space flow is like an incompressible fluid \citep{Strauch2009,Eastman2015}
\begin{equation}
\frac{d\rho}{dt}=\frac{\partial\rho}{\partial t}+\sum_{i}\left(\frac{\partial\rho}{\partial q_{i}}\frac{dq_{i}}{dt}+\frac{\partial\rho}{\partial p_{i}}\frac{dp_{i}}{dt}\right)=0 .
\end{equation}
This implies that the probability density remains constant and the
physical system occupies the same volume in phase space at all times.
In other words, the number of microstates neither increases, nor decreases,
which is a manifestation of the fact that dynamic trajectories neither
bifurcate, nor merge. Whether we know or do not know the exact microstate
of the classical physical system, has no bearing on the inability
of classical systems to make choices.

\paragraph{Chaotic dynamics}
Nonlinear dynamics in multipartite classical systems can lead to deterministic chaos \citep{Scott2007}. Chaotic behavior is characterized with extreme sensitivity to small perturbations, manifestation of irregular orbits that explore the entire phase space, and separation of infinitesimally close orbits at an exponentially fast rate, which is quantifiable in terms of the Lyapunov exponent \citep{Ramaswamy1984}. Taken together, these features of chaotic systems establish a short-term predictability time window during which the \emph{actual orbit} of the system does not deviate significantly from the \emph{predicted orbit}. For longer times, however, the dynamics of chaotic systems becomes practically unpredictable due to exponential amplification of tiny errors in the empirical measurement of the initial state of the system \citep{Lorenz1963a,Lorenz1963b}. This kind of \emph{effective unpredictability}, which is due to our lack of knowledge, skills or technologies to predict the deterministic orbits far enough into the future \citep{Hawking2010}, has no bearing on the fact that classical deterministic systems are unable to make choices \citep{Georgiev2017}. In fact, a classical deterministic universe has no internal source of perturbations, which means that under the phrase ``small perturbations'' the classical physicist hides any errors that arise due to the simplifying assumption that the physical system of interest is closed and the rest of the universe can be ignored. Alternatively, working with a finite numerical precision while solving a system of differential equations \citep{Hirsch2013} could be viewed as introducing fictitious perturbations at each computational step where numerical rounding is employed. Since free will is the inherent capacity of physical systems to make genuine choices, it cannot be rescued by mere consideration of instability and chaos in complex classical systems. Mathematical modeling of genuine choices requires the introduction of physical propensities and indeterminism.

\paragraph{Incompatibilism}
The philosophical stance that free will is incompatible with physical
determinism is referred to as incompatibilism. Deducing incompatibilism,
however, is only the starting point of any serious investigation of
the problem of free will. Before the discovery of quantum mechanics in 1920s,
all leading physical theories, including Newtonian, Lagrangian, and Hamiltonian mechanics, as well as Maxwell's electrodynamics, were deterministic. This meant that any classical physical theory of consciousness was forced
to accept determinism and abandon free will. Without free will there
can be no moral responsibility, at least no more than a stone is morally
responsible for breaking someone's leg or a car is morally responsible
for not moving without petrol fuel \citep{Russell2005}.
Under ordinary circumstances, arriving at a contradiction with our own introspective testimony of free will (for a list of such contradictions see \ref{app}) would be viewed as an indication to reject the faulty premise, which happens to be classical physics.
Yet, there are still some philosophers and/or neuroscientists who reject free will \citep{Dennett2004,Harris2012,Wegner2003} based on the outdated belief in determinism resulting from Hamilton's equations.
Before 1920s, classical physics was essentially all of physics, which meant that
to defend free will one had to abandon the physicalism altogether.
Thus, it is not surprising that a number of prominent 19th century
philosophers chose physicalism over free will \citep{Locke1828,Schopenhauer1841}.
With the advent of modern quantum mechanics, however, we now know
that classical mechanics in inadequate to describe the physical world
and physical reality is governed by indeterministic quantum physical laws.
Modern quantum physics no longer clashes with the existence of free
will and supports the possibility of genuine choice making.

\section{Quantum indeterminism and free will}
\label{sec:Quantum}

Having quantified the amount of free will $\mathcal{F}$, which is intimately
linked to the fundamental physical laws that govern the dynamics of
physical systems, we can use \eqref{eq:fw} as a tool to demonstrate
that free will may evolve through natural selection in physical agents
that inhabit a quantum indeterministic world. The important thing
to note is that in an indeterministic world, the physical laws may
predict different probability distributions for different situations.
In the presence of genuine bifurcations of the dynamic trajectories
for future courses of action, the choices of the physical
agents may have important consequences about whether the agent will
have the same abundance of future courses of action to choose from.
For example, based on the outcome of a past choice some of the previously
available trajectories may now obtain zero probability of future occurrence.
To make these ideas more tangible, first we will briefly explain how
the Schr\"{o}dinger equation and the Born rule are used to predict the
probabilities for different outcomes in quantum measurements. Then,
we will illustrate how quantum theory supports the full range of actions,
from completely unbiased to completely biased, using the famous Stern--Gerlach
experiment implementing alternative (incompatible) quantum measurements
of the spin components of a single qubit.

\paragraph{Schr\"{o}dinger equation}
The main quantum physical law that governs
\emph{what exists} and \emph{how it evolves in time} is given by the
Schr\"{o}dinger equation \citep{Schrodinger1928,Hayashi2015} 
\begin{equation}
\imath\hbar\frac{\partial}{\partial t}|\Psi(\mathbf{r},t)\rangle=\hat{H}\,|\Psi(\mathbf{r},t)\rangle ,
\end{equation}
where $\imath=\sqrt{-1}$ is the imaginary unit, $\hbar$ is the reduced Planck constant,
$\frac{\partial}{\partial t}$
is the partial derivative operator with respect to time, $|\Psi(\mathbf{r},t)\rangle$
is the quantum state vector, $\mathbf{r}=\left(x,y,z\right)$ is the
vector of position coordinates, $t$ is time, and $\hat{H}$ is the
Hamiltonian operator corresponding to the total energy of the quantum
system \citep{Georgiev2017,Georgiev2020a,Georgiev2020b,Georgiev2021}.

The quantum state vector $|\Psi(\mathbf{r},t)\rangle$ of the physical
system represents a continuous distribution of quantum probability amplitudes,
which exhibit wave-like properties in 3-dimensional space \citep{Georgiev2018a,Georgiev2021b}.
For example, the linearity of the Schr\"{o}dinger equation implies that any two solutions
$|\Psi_{1}(\mathbf{r},t)\rangle$ and $|\Psi_{2}(\mathbf{r},t)\rangle$
can interfere with each other in the form of a linear quantum superposition
\begin{equation}
|\Psi_{s}(\mathbf{r},t)\rangle=a_{1}|\Psi_{1}(\mathbf{r},t)\rangle+a_{2}|\Psi_{2}(\mathbf{r},t)\rangle ,
\end{equation}
where $a_{1}$ and $a_{2}$ are complex coefficients. Due to the principle
of quantum superposition, the quantum state behaves like a vector
in an abstract Hilbert space. For an $n$-dimensional Hilbert space
$\mathcal{H}$, the quantum state is an $n\times1$ column vector
called a \emph{ket} \citep{Dirac1939,Dirac1967,Bhaumik2019}
\begin{equation}
|\Psi(\mathbf{r},t)\rangle=\left(\begin{array}{c}
a_{1}\\
a_{2}\\
\vdots\\
a_{n}
\end{array}\right) .
\end{equation}
The complex conjugate transpose of the ket is a $1\times n$ row vector
called a \emph{bra} 
\begin{equation}
\langle\Psi(\mathbf{r},t)|=\left(\begin{array}{cccc}
a_{1}^{*} & a_{2}^{*} & \ldots & a_{n}^{*}\end{array}\right)
\end{equation}
residing in a dual Hilbert space $\mathcal{H}^{*}$ \citep{Berezansky1996}.
Normalization of physical probabilities requires that the quantum states
have a unit norm
\begin{equation}
\langle\Psi(\mathbf{r},t)|\Psi(\mathbf{r},t)\rangle=\sum_{n}a_{n}^{*}a_{n}=\sum_{n}|a_{n}|^{2}=1 .
\end{equation}

\paragraph{Born rule}
Quantum indeterminism is manifested in the act of
quantum measurement, which is governed by the Born rule \citep{Born1955,Busch2016}.
\emph{What can be measured} as observable physical quantities are
the \emph{eigenvalues} $\lambda_{1},\lambda_{2},\ldots,\lambda_{n}$
of some \emph{quantum operator} (also called \emph{quantum
observable}) $\hat{A}$, which is represented by an $n\times n$ matrix
and indicated with the hat symbol \citep{Dirac1967,Fayngold2013,Susskind2014,Hayashi2015}.
The quantum observable $\hat{A}$ may operate upon any input quantum
state and return another output quantum state \citep{Holevo2001}.
However, of special physical significance is the set of eigenvectors
$|\Phi_{1}\rangle,|\Phi_{2}\rangle,\ldots,|\Phi_{n}\rangle$ of $\hat{A}$
such that the action of $\hat{A}$ on an eigenvector $|\Phi_{n}\rangle$
returns the same eigenvector $|\Phi_{n}\rangle$ multiplied by the
corresponding eigenvalue $\lambda_{n}$, namely, $\hat{A}|\Phi_{n}\rangle=\lambda_{n}|\Phi_{n}\rangle$
\citep{Strang2016}. The eigenvectors and eigenvalues of quantum observables
allow a spectral decomposition in the form
\begin{equation}
\hat{A}=\sum_{n}\lambda_{n}|\Phi_{n}\rangle\langle\Phi_{n}|=\left(\begin{array}{cccc}
\lambda_{1} & 0 & 0 & 0\\
0 & \lambda_{2} & 0 & 0\\
0 & 0 & \ddots & 0\\
0 & 0 & 0 & \lambda_{n}
\end{array}\right) ,
\end{equation}
where $|\Phi_{1}\rangle\langle\Phi_{1}|$, $|\Phi_{2}\rangle\langle\Phi_{2}|$,
$\ldots$, $|\Phi_{n}\rangle\langle\Phi_{n}|$ are the individual
projection operators onto the rays $|\Phi_{1}\rangle,|\Phi_{2}\rangle,\ldots,|\Phi_{n}\rangle$
in the Hilbert space $\mathcal{H}$.

The Born rule states that if the quantum observable $\hat{A}$ is
measured on a quantum physical system in state $|\Psi\rangle$, then
the expected value $\bar{A}$ (determined as weighted average over all observable outcomes) is
\begin{equation}
\bar{A}=\langle\Psi|\hat{A}|\Psi\rangle=\sum_{n}\lambda_{n}\langle\Psi|\Phi_{n}\rangle\langle\Phi_{n}|\Psi\rangle=\sum_{n}\lambda_{n}a_{n}^{*}a_{n}=\sum_{n}|a_{n}|^{2}\lambda_{n} , \label{eq:Born}
\end{equation}
where $a_{n}=\langle\Phi_{n}|\Psi\rangle$ is the projected quantum
probability amplitude from the state $|\Psi\rangle$ onto the state
$|\Phi_{n}\rangle$, $a_{n}^{*}$ is the complex conjugate of $a_{n}$,
and $|a_{n}|^{2}$ is the quantum probability for the measuring device
to register the observable outcome $\lambda_{n}$.

\paragraph{Alternative measurements of the quantum spin of a single qubit}
Different quantum probability distributions arise for alternative
(incompatible) quantum measurements as illustrated by the Stern--Gerlach
experiment with silver atoms \citep{Gerlach1922,Cruz-Barrios2000,Potel2005}. Each silver atom
is a simple two-level quantum system (qubit) that exhibits only two
possible values $\pm\frac{1}{2}$ (in $\hbar$ units) of the quantum
spin observable $\hat{S}$.

The quantum spin of a qubit can point along an arbitrary $u$-axis
inside the real 3-dimensional space. Expressed in spherical coordinates
$(r,\theta,\varphi)$, with the normalization condition $r=1$ taken
into account, the spin observable $\hat{S}_{u}$ along the $u$-axis
can be written in terms of the Pauli spin matrices $\hat{\sigma}_{x}$, $\hat{\sigma}_{y}$ and $\hat{\sigma}_{z}$ as
\begin{equation}
\hat{S}_{u}=\sin\theta\cos\varphi\,\hat{\sigma}_{x}+\sin\theta\sin\varphi\,\hat{\sigma}_{y}+\cos\theta\,\hat{\sigma}_{z} .
\end{equation}

If we perform experiments with a qubit that is initially
prepared in an eigenstate of the $\hat{S}_{z}$ observable, it will
be convenient to express all quantum states and observables in the
$|\uparrow_{z}\rangle,|\downarrow_{z}\rangle$ basis as follows
\begin{equation}
|\uparrow_{z}\rangle=\left(\begin{array}{c}
1\\
0
\end{array}\right),\quad|\downarrow_{z}\rangle=\left(\begin{array}{c}
0\\
1
\end{array}\right),
\end{equation}
\begin{equation}
\hat{\sigma}_{x}=\left(\begin{array}{cc}
0 & 1\\
1 & 0
\end{array}\right),\quad\hat{\sigma}_{y}=\left(\begin{array}{cc}
0 & -\imath\\
\imath & 0
\end{array}\right),\quad\hat{\sigma}_{z}=\left(\begin{array}{cc}
1 & 0\\
0 & -1
\end{array}\right),
\end{equation}
\begin{equation}
\hat{S}_{u}=\left(\begin{array}{cc}
\cos\theta & e^{-\imath\varphi}\sin\theta\\
e^{\imath\varphi}\sin\theta & -\cos\theta
\end{array}\right) ,
\end{equation}
where $\theta$ is the polar angle and $\varphi$ is the azimuthal
angle of the $u$-axis. The eigenvectors of $\hat{S}_{u}$ are
\begin{eqnarray}
|\uparrow_{u}\rangle & = & \cos\left(\frac{\theta}{2}\right)|\uparrow_{z}\rangle+\sin\left(\frac{\theta}{2}\right)e^{\imath\varphi}|\downarrow_{z}\rangle ,\\
|\downarrow_{u}\rangle & = & -\sin\left(\frac{\theta}{2}\right)|\uparrow_{z}\rangle+\cos\left(\frac{\theta}{2}\right)e^{\imath\varphi}|\downarrow_{z}\rangle ,
\end{eqnarray}
with corresponding eigenvalues of $+\frac{1}{2}$ for $|\uparrow_{u}\rangle$
and $-\frac{1}{2}$ for $|\downarrow_{u}\rangle$.

Substitution of different values for the polar and azimuthal angles
gives the eigenvectors for the Pauli matrices: the eigenvectors $|\uparrow_{x}\rangle,|\downarrow_{x}\rangle$
of $\hat{S}_{x}=\hat{\sigma}_{x}$ are obtained for $\theta=\frac{\pi}{2}$
and $\varphi=0$, the eigenvectors $|\uparrow_{y}\rangle,|\downarrow_{y}\rangle$
of $\hat{S}_{y}=\hat{\sigma}_{y}$ are obtained for $\theta=\frac{\pi}{2}$
and $\varphi=\frac{\pi}{2}$, and the eigenvectors $|\uparrow_{z}\rangle,|\downarrow_{z}\rangle$
of $\hat{S}_{z}=\hat{\sigma}_{z}$ are obtained for $\theta=0$ and
$\varphi=0$.

\begin{figure}[t]
\begin{centering}
\includegraphics[width=160mm]{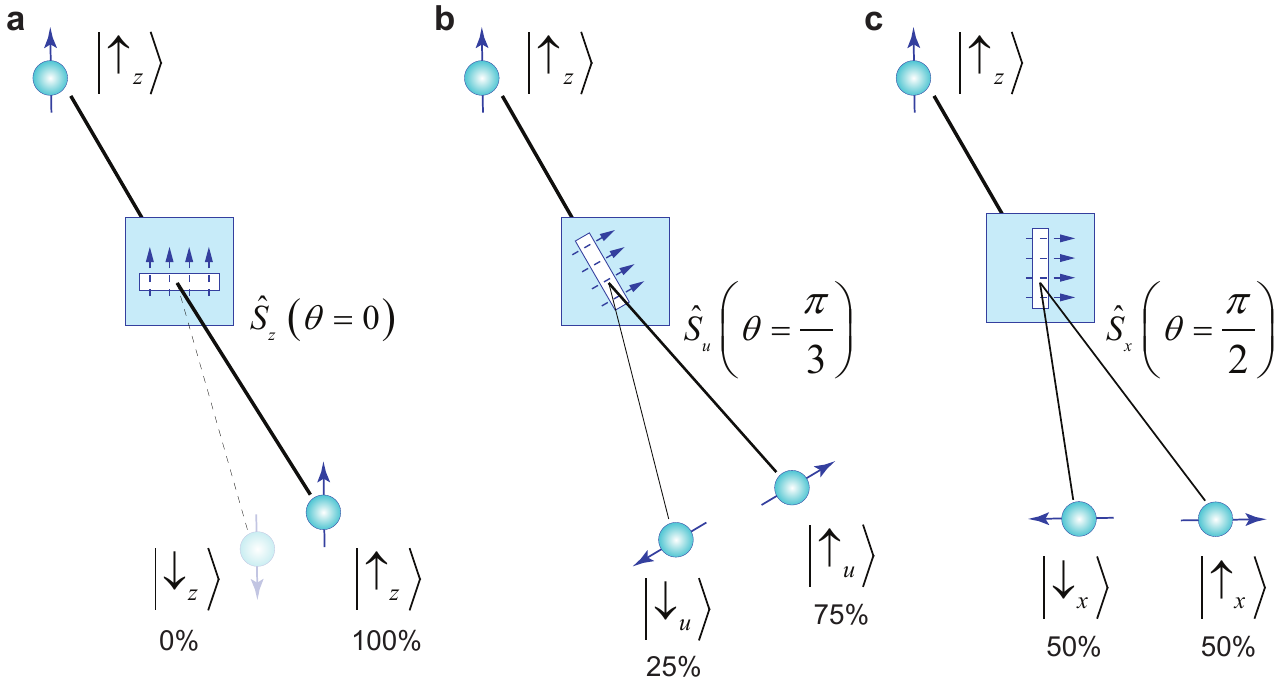}
\par\end{centering}
\caption{\label{fig:2}Stern--Gerlach experiment with a silver atom (qubit) whose initial
state is $|\Psi\rangle=|\uparrow_{z}\rangle$, which passes through
inhomogeneous magnetic field rotated at a polar angle $\theta$. Depending
on which quantum observable is measured, the outcomes can vary from
completely deterministic to completely indeterministic: (a) If $\hat{S}_{z}(\theta=0)$
is measured, the observable outcome $|\uparrow_{z}\rangle$ occurs
with probability of 100\% and $|\downarrow_{z}\rangle$ with probability
of 0\%. (b) If $\hat{S}_{u}(\theta=\frac{\pi}{3})$ is measured, the
observable outcome $|\uparrow_{u}\rangle$ occurs with probability
of 75\% and $|\downarrow_{u}\rangle$ with probability of 25\%. (c)
If $\hat{S}_{x}(\theta=\frac{\pi}{2})$ is measured, the observable
outcome $|\uparrow_{x}\rangle$ occurs with probability of 50\% and
$|\downarrow_{x}\rangle$ with probability of 50\%.}
\end{figure}

Now we are ready to compute the quantum probability distributions
for a qubit whose initial state is $|\Psi\rangle=|\uparrow_{z}\rangle$,
which passes through a Stern--Gerlach apparatus measuring one of three
alternative (incompatible) spin observables: $\hat{S}_{z}\left(\theta=0\right)$,
$\hat{S}_{u}\left(\theta=\frac{\pi}{3}\right)$ or $\hat{S}_{x}\left(\theta=\frac{\pi}{2}\right)$,
where without loss of generality we have set $\varphi=0$.
In the case when $\hat{S}_{z}$ is measured (Figure~\ref{fig:2}a), the
observable outcome $|\uparrow_{z}\rangle$ occurs with probability
of 100\% and $|\downarrow_{z}\rangle$ with probability of 0\%.
In other words, because the initial state $|\Psi\rangle=|\uparrow_{z}\rangle$
is an eigenvector of the measured observable $\hat{S}_{z}$, the outcome
is absolutely certain and deterministic. In fact, the capacity to
produce a deterministic outcome upon measurement of a quantum observable
is a physical way of defining what an eigenvector of that quantum
observable is.
In the case when $\hat{S}_{u}\left(\theta=\frac{\pi}{3}\right)$ is measured (Figure~\ref{fig:2}b),
the observable outcome $|\uparrow_{u}\rangle$ occurs with probability
of 75\% and $|\downarrow_{u}\rangle$ with probability of 25\%. The
observable outcomes are indeterministic but biased. Completely unbiased
indeterministic outcomes are obtained when $\hat{S}_{x}$ is measured
(Figure~\ref{fig:2}c), namely, the observable outcome $|\uparrow_{x}\rangle$
occurs with probability of 50\% and $|\downarrow_{x}\rangle$ with
probability of 50\%. Taken together, these alternative experimental
conditions illustrate a crucial fact, namely, the act of quantum measurement
supports the whole range of probability distributions, from completely
deterministic to completely indeterministic, depending on the measured physical observable parametrized by $\theta$.
In other words, stating
that quantum physics is indeterministic should be understood in the
sense that quantum physics admits indeterminism as a possibility without
excluding determinism as a special case of biased quantum
probability distribution.

Noteworthy, quantum probabilities remain invariant with respect to
interchange of the initial and final states during quantum measurement.
For example, consider an initial state $|\Psi\rangle$ and final state
$|\Phi_{n}\rangle$. The probability $P_{n}$ to observe the transition
$|\Psi\rangle\to|\Phi_{n}\rangle$ in the act of quantum measurement
is given by the expectation value of the projection operator $|\Phi_{n}\rangle\langle\Phi_{n}|$
for the initial state $|\Psi\rangle$ according to the Born rule
\begin{equation}
P_{n}=\langle\Psi|\Phi_{n}\rangle\langle\Phi_{n}|\Psi\rangle=a_{n}^{*}a_{n}=|a_{n}|^{2} .
\end{equation}
Because the multiplication of complex quantum probability amplitudes
is commutative (namely, the order of multiplication does not matter, $a_{n}^{*}a_{n}=a_{n}a_{n}^{*}$),
$P_{n}$ is also the probability to observe the converse transition
$|\Phi_{n}\rangle\to|\Psi\rangle$ given by the expectation value
of the projection operator $|\Psi\rangle\langle\Psi|$ for the initial
state $|\Phi_{n}\rangle$ as follows 
\begin{equation}
P_{n}=\langle\Phi_{n}|\Psi\rangle\langle\Psi|\Phi_{n}\rangle=a_{n}a_{n}^{*}=|a_{n}|^{2} .
\end{equation}
Thus, we have a second physical way to obtain alternative quantum probability
distributions, namely, we can measure the same quantum observable $\hat{S}_{z}$
for different initial quantum states $|\uparrow_{z}\rangle$, $|\uparrow_{u}\rangle$
or $|\uparrow_{x}\rangle$. We already know that if $|\uparrow_{z}\rangle$
is the initial state, the observable outcome $|\uparrow_{z}\rangle$
occurs with probability of~100\% and $|\downarrow_{z}\rangle$ with
probability of~0\%. In the case when $|\uparrow_{u}\rangle$ is the
initial state, the observable outcome $|\uparrow_{z}\rangle$ occurs
with probability of~75\% and $|\downarrow_{z}\rangle$ with probability
of~25\%. Completely unbiased quantum probability distribution is
obtained when $|\uparrow_{x}\rangle$ is the initial state, where
the observable outcome $|\uparrow_{z}\rangle$ occurs with probability
of~50\% and $|\downarrow_{z}\rangle$ with probability of~50\%.

Equipped with the precise understanding of the Born rule~\eqref{eq:Born},
and knowing the two main physical ways for production of biased probability distributions in the process of quantum measurement,
we are now ready to tackle the problem of how free will is able to evolve
in biological systems by varying the parameter $\theta$.

\section{Neurophysiological mechanisms of free will}
\label{sec:Neurophysiology}

Free will is often viewed as a binary property: the capacity of a
conscious agent to do otherwise is either true or false by virtue
of the physical laws. For centuries philosophers felt no urgent need
to quantify free will in the presence of biased choices, because the
verdict in classical physics was clear, namely, physical determinism
forbids free will and there is no point in quantifying something that
is an illusion. In the modern age of advanced quantum technologies,
however, we know that the physical reality is nonclassical and quantum
indeterminism endows quantum physical systems with inherent propensity
to make genuine choices, thereby manifesting varying amount of free
will depending on how biased those choices are.

Without the existence of genuinely quantum physical substrates in the brain, the conscious agents would not have been capable of harnessing quantum effects to support intelligence, thinking and decision making \citep{Matsuno2000,Matsuno2006,Melkikh2019,Georgiev2017}.
The interaction between the constituent quantum particles of the nervous
system of an organism and the surrounding environment constitutes
a quantum measurement in the act of which different observable outcomes
could occur with different probabilities.
Similarly to the Stern--Gerlach experiment discussed above, the environment would be able to measure
alternative (incompatible) quantum observables of neurons, which would
comprise a set of functional conformations of neural biomolecules
with catalytic activities. For example, quantum chemistry research
supports dynamic quantum effects such as quantum tunneling
in the gating of voltage-gated ion channels \citep{Chancey1992,Vaziri2010,Kariev2019,Kariev2021}
or in the zipping of SNARE proteins during neurotransmitter release \citep{Georgiev2018b,Georgiev2020b}.

\begin{figure}[t]
\begin{centering}
\includegraphics[width=154mm]{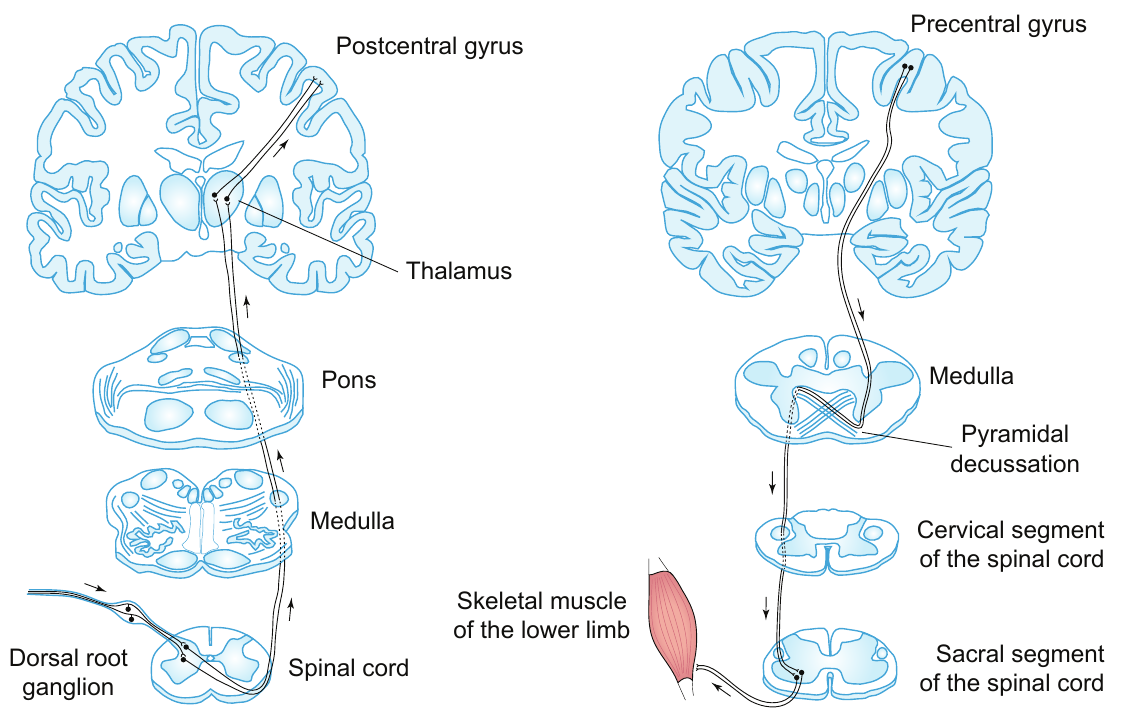}
\par\end{centering}
\caption{\label{fig:3}Classical communication through electric signals between
the brain cortex and the body. The somatosensory pathway (left) delivers
sensory information from the body to the somatosensory cortex in the
postcentral gyrus, whereas the somatomotor pathway (right) delivers
motor information from the motor cortex in the precentral gyrus to
the body muscles. Deterministic release of synaptic vesicles lacks free will but ensures error-free perception of the surrounding world and guarantees reliable execution of fight-or-flight behavioral responses.
The spinal cord segments, medulla and pons are represented
with their transversal sections, whereas thalamus and cortex are shown
in frontal slice. Modified from \cite{Georgiev2017}.}
\end{figure}

\paragraph{Deterministic synaptic neurotransmission in sensory or motor pathways}
Electric spikes (action potentials) propagating along the neuronal
projections inside the neural network are the main carriers of physical
information and consume about half of the energy utilized for supporting
clinical consciousness by the brain cortex \citep{Georgiev2020e}.
Sensory information is encoded in the form of electric spikes by the
peripheral sensory organs after which it is reliably transmitted across
the synapses of the sensory pathways through the thalamus toward the
brain cortex where it is consciously experienced \citep{Kim2013,Singer2007,Glowatzki2002,Magistretti2015}.
Somatomotor information from the motor cortex is outputted again in
the form of electric signals that are reliably transmitted across
the synapses of the somatomotor pathways through the anterior horn
of the spinal cord toward the muscles \citep{Kuno1971}. To achieve
reliable signal transmission, the chemical synapses in these pathways
operate in a deterministic fashion employing multivesicular release
upon depolarization of the presynaptic axonal boutons \citep{Rudolph2015,Pulido2017}.
The exocytotic release of neurotransmitter molecules from multiple
synaptic vesicles then generates large postsynaptic currents in the
target neuron \citep{Rudolph2015}. The reliable deterministic transmission
of information between the brain cortex and the body (Figure~\ref{fig:3})
is of paramount importance for the survival of the organism through
the execution of fight-or-flight responses.

Synaptic vesicle release is a physical process that can be represented
by a particular quantum observable $\hat{A}=1|1\rangle\langle1|+0|0\rangle\langle0|$,
which can be spectrally decomposed into two coarse-grained eigenvectors
denoted as $|1\rangle$ with eigenvalue $1$ for release of at least
one synaptic vesicle and $|0\rangle$ with eigenvalue $0$ for no
synaptic vesicles released \citep{Beck1992}. Formally, the mathematical structure of
$\hat{A}$ is the same as the spin observable projection operator
$|\uparrow_{z}\rangle\langle\uparrow_{z}|$. Under normal physiological
conditions, the probability of release of at least one synaptic vesicle
in extracortical synapses in the sensory/somatomotor pathways is 100\%.
This complete determinism implies that the quantum state of the extracortical
synapse is $|\Psi\rangle=|1\rangle$ when the quantum measurement
is performed, $\langle\Psi|\hat{A}|\Psi\rangle=1$, hence no free
will is manifested.

\begin{figure}[t]
\begin{centering}
\includegraphics[width=140mm]{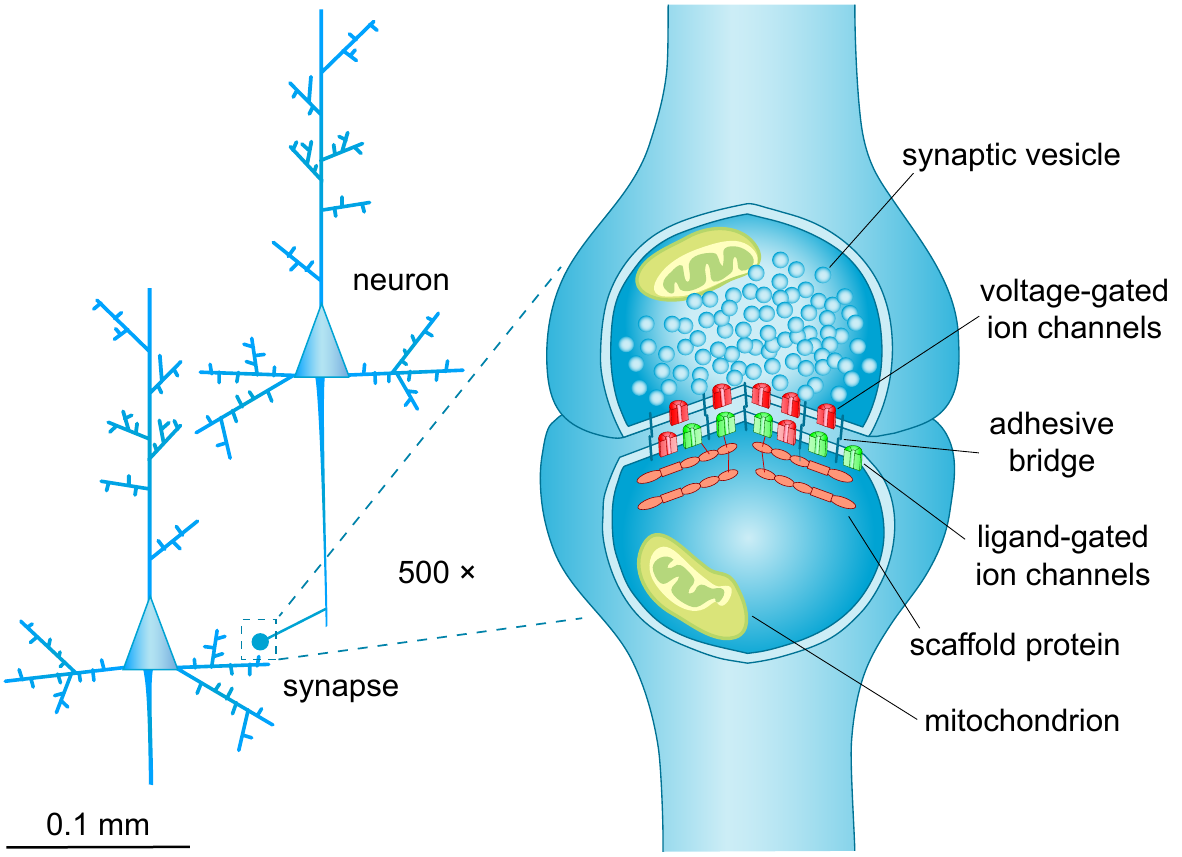}
\par\end{centering}
\caption{\label{fig:4}Excitatory synaptic contact between pyramidal cortical
neurons. The presynaptic axonal bouton has a pool of synaptic vesicles
that contain neurotransmitter. During an electric spike, the activation
of presynaptic voltage-gated calcium channels initiates Ca\protect\textsuperscript{2+}
influx at the active zone, which may trigger fusion of a single synaptic
vesicle with the plasma membrane.
Varying probability of release provides different amounts of free will exercised in the synapses of the brain cortex. This inherent indeterminism supports higher cognitive decisions that are less predictable by adversaries and may enhance the survival of organisms.
Following the successful exocytosis of a synaptic vesicle, the released neurotransmitter acts
on postsynaptic receptors to induce postsynaptic electric currents
in the target neuron. Structural support for the synapse is provided
by adhesive bridges and scaffold proteins, whereas mitochondria ensure
robust energy supply for synaptic neurotransmission. Modified from
\cite{Georgiev2017}.}
\end{figure}

\paragraph{Indeterministic synaptic neurotransmission in the brain cortex}
Individual synapses of pyramidal neurons inside the neocortex and
the hippocampus operate in an indeterministic fashion releasing
either a single synaptic vesicle or none \citep{Beck1992,Arancio1994,Matveev2000,Hanse2001},
which is consistent with possible direct involvement in the neural
mechanisms supporting human consciousness and free will. Each cortical
synapse (Figure~\ref{fig:4}) appears to possess only one functional
release site at a given time \citep{Stevens1995} such that the probability
for exocytosis is $0.35\pm0.23$ per axonal spike \citep{Dobrunz1997}.
Membrane-bound SNARE protein complexes zip into four-$\alpha$-helix
bundles that drive exocytosis by merging the synaptic vesicles with
the presynaptic plasma membrane \citep{Weber1998,Han2017}. Quantum
vibrational excitons propagating along the protein $\alpha$-helices
might be instrumental in producing indeterministic physical outcomes
with the use of quantum tunneling through massive barriers that are
imposed by external protein clamps \citep{Georgiev2019b,Georgiev2019c,Georgiev2020c,Georgiev2020d,Georgiev2020b}.
In the absence of electric spikes, each SNARE protein complex is clamped
by the Ca\textsuperscript{2+} sensor protein synaptotagmin-1 to prevent
spontaneous release from resting neurons \citep{Chapman2002,Zhou2015,Zhou2017}.
Only when the presynaptic buttons are depolarized, the activation
of voltage-gated calcium channels leads to entry of Ca\textsuperscript{2+}
ions that assist in releasing the synaptotagmin-1 clamping action,
enhancing the probability of quantum tunneling of the vibrational
exciton through the barrier and zipping the SNARE protein complex \citep{Georgiev2018b}.
Cortical pyramidal neurons form on average 7500~synapses onto target neurons
\citep{Braitenberg1998}, where each electrically excited axonal bouton
of a cortical pyramidal neuron is almost twice more likely to fail
than succeed in releasing neurotransmitter \citep{Georgiev2018b,Georgiev2020e}.
This means that the quantum state of the extracortical synapse is
$|\Psi\rangle=\sqrt{0.35}|1\rangle+\sqrt{0.65}|0\rangle$ when the
quantum measurement is performed, $\langle\Psi|\hat{A}|\Psi\rangle=0.35$,
hence the amount of free will manifested through synaptic vesicle
release is $\approx0.934$ bits per synapse or $\approx7005$ bits
per cortical pyramidal neuron. The human brain cortex contains $1.634\times10^{10}$
neurons \citep{Azevedo2009} of which at least 70\% are excitatory
pyramidal neurons \citep{Nieuwenhuys1994}. If all cortical pyramidal
neurons fire once, the amount of free will exercised would be over
80\,terabits ($8\times10^{13}$\,bits). In other words, the expected
information gain from learning which synapses are active and which
synapses remain silent during an electric firing of all cortical pyramidal
neurons is sufficient to exhaust the memory of a modern 10\,terabyte
hard drive (1\,byte = 8\,bits). The energy power of 4.427 W consumed
by the human brain cortex suffices to sustain an average firing frequency
of 9.6 Hz \citep{Georgiev2020e}. Thus, the overall amount of free
will provided by cortical synaptic activity is over 96 terabytes per
second.

Taking stock of matters so far, we see that animal evolution through
natural selection could have indeed optimized the amount of free will
manifested by cortical or extracortical synapses in the nervous system.
In order to receive reliable sensory information, the animal sensory
organs and the sensory pathways need to operate deterministically
with zero free will. Otherwise, the animal will receive various artifacts
(errors) in the sensory picture of the world, which will result in
inappropriate responses, possible injuries and ultimately death. Similarly,
the somatomotor information outputted towards the muscles needs to
be reliably transmitted for the execution of appropriate fight-or-flight
responses.

\paragraph{Importance of free will in the evolution of prey--predator relationship}
Consider a wild gazelle hunted by a pride of lions in the savanna.
In order to outrun the predators, the motor cortex of the gazelle
needs to control reliably the contraction of skeletal muscles, which
also requires deterministic operation and lack of free will along
the synapses of the somatomotor pathway. Failure of the gazelle to
execute flawlessly the flight response \citep{Walther1969} would lead
to demise. The perfect communication between the brain cortex and
the body, however, would not be very helpful for the prey animal if
the overall behavior were deterministic and easily predictable by
the predators. The hunting pride of lions usually attempts to ambush
and encircle the prey. If the gazelle were running in a straight direction,
its future path would be easily predictable by the lions and they
would easily intercept it. That is why a certain amount of indeterministic
performance of the brain cortex becomes an invaluable asset that can
be further evolved to an optimal level through natural selection.
Using its inherent free will, the gazelle is capable to unpredictably
jump to the left or to the right, thereby confusing the lions and
escaping through gaps in the circle left unattended by the lions.
This is how enhancing the amount of free will possessed by the brain
cortex may affect the animal behavior and have a survival value for
the organism.

\paragraph{Importance of free will for human creativity}
Divergent thinking characterizes creative thought during which are combined concepts and ideas that were previously thought to be unrelated \citep{Hudson1974,Cardoso2009,Runco2020}. Free will provides an inherent probabilistic mechanism for ideation of possible creative solutions, which are then critically assessed and their merits evaluated in a convergent process of elimination \citep{Georgiev2018c}. Successful creative solutions and human discoveries can then be transmitted in the form of art or passed down culturally from generation to generation \citep{MorrissKay2010,Hodgson2015}.

\section{Free will and moral responsibility}
\label{sec:Moral}

Free will is intimately connected with ethics and moral responsibility
\citep{Kane1994,Kane1996b,Kane2000a,Kane2016}. Conscious agents with
free will are able to choose whether to do an action or not. In either
way, action or inaction, the agents are morally responsible for the
consequences of their choices, including any opportunities missed
from the actions that they have chosen not to do. Moral responsibility,
however, is contingent on the freedom of choice and the lack of
external coercion. The amount of external coercion $\mathcal{C}$
could be quantified using the \emph{Kullback--Leibler divergence}
\eqref{eq:kullback} from an initial distribution $P_{i}(x_{k})$
characterizing the inherent desires of the agent to a final distribution
$P_{f}(x_{k})$ characterizing the available choices in the presence
of coercive force 
\begin{equation}
\mathcal{C}=\sum_{k}P_{f}(x_{k})\log_{2}\left[\frac{P_{f}(x_{k})}{P_{i}(x_{k})}\right] . \label{eq:coercion}
\end{equation}
In the case when a single course of action $x_{k}$ is forced onto
an agent, $P_{f}(x_{k})=1$, the amount of coercion reduces to the
surprisal \eqref{eq:surprisal} resulting from occurrence of the individual
outcome $x_{k}$ in the absence of coercion
\begin{equation}
\mathcal{C}=-\log_{2}P_{i}(x_{k}) .
\end{equation}
In other words, there is no coercion, $\mathcal{C}=0$, if the agent
would have chosen the outcome $x_{k}$ anyway, $P_{i}(x_{k})=1$. The
coercion is $\mathcal{C}=n$ bits for $P_{i}(x_{k})=(\frac{1}{2})^{n}$.
In the limiting case when the agent has no desire to choose the outcome
$x_{k}$ at all, namely, $P_{i}(x_{k})=0$, the coercion becomes infinitely
large, $\mathcal{C}=\infty$.

In classical deterministic physics, free will is impossible and the
concept of coercion becomes meaningless. If the universe is a clockwork
mechanism and humans are nothing but string puppets reacting to their
physical circumstances, then nobody is accountable for anything. In
the first half of 20th century, the famous trial lawyer Clarence Darrow
successfully defended murderers from receiving the death penalty using
the \emph{domino theory of moral nonresponsibility}, namely, if someone's
actions are always consequences of what others have done to him or
her, then no one is responsible for anything he or she does \citep{Darrow1927,Darrow1932}.
Indeed, we do not blame or punish falling stones for breaking someone's
leg or static cars for not moving without petrol fuel \citep{Russell2005}.

In a quantum indeterministic world, however, free will is a valuable
asset because it allows everyone to be able to choose his or her own
purpose and meaning of life \citep{Georgiev2017}. The free choices
do not imply physical lawlessness. Instead, fundamental quantum physical
laws including the Schr\"{o}dinger equation and the Born rule determine
the available future alternatives from which we are allowed to choose
and how biased the probabilities for those different choices are.
Because the acts of choosing are executed during quantum measurements
performed by the surrounding environment upon the quantum state of
the neural network, the organism is able to form memories of past
choices and adapt the internal quantum dynamics so that the quantum
probabilities for release of a synaptic vesicle per action potential
can vary widely between cortical axonal buttons \citep{Dobrunz1997,Trommershauser2003,Branco2009,Korber2016,Taschenberger2016,Volgushev2004}.
Adjustment of synaptic vesicle release probabilities could be easily
achieved by varying the potential energy barriers for quantum tunneling
of vibrational excitons propagating along the $\alpha$-helices of
SNARE proteins \citep{Georgiev2017,Georgiev2018b,Georgiev2019c,Georgiev2020b}.
Changing the quantum probability distributions for synaptic release
modifies the amount of free will manifested by a conscious agent at
different stages in life.

\paragraph{Drug addiction, free will and moral responsibility}
Narcotic drugs are able to elicit an initial surge of pleasurable sensation followed
by a compulsive drive for self-administration of the drug \citep{Volkow2004,Luscher2020}
mostly to avoid undesirable side effects of drug withdrawal \citep{Koob2020}
such as restlessness, irritability, insomnia, muscle and bone pain,
muscle spasms, kicking movements of the legs, abdominal pain, diarrhea,
vomiting, chills or cold flashes with goose bumps. If a drug addict
is going to self-administer the drug with absolute certainty, then
it would appear that there is no free will manifested. This, however,
does not imply that the drug addict bears no moral responsibility
for his or her actions, because one could rewind back the time until
the first exposure to the addictive drug. At this point in time, if
the subject was able to do otherwise but did not choose to do so (e.g., due to curiosity of trying what is it like to experience the surge
of pleasurable sensation elicited by the narcotic drug), then the
subject is fully responsible for his actions as he or she could have
exercised at least $\mathcal{F}=1$ bit of free will. Alternatively,
if the subject had no intention to try any drugs but the first
drug administration was forced by someone else (e.g., drink spiking without subject's knowledge or permission, criminal
intravenous injection, etc.), then the subject is not responsible
for becoming an addict as he or she was subjected to an infinitely
large coercion, $\mathcal{C}=\infty$. This clearly shows that because
the amount of free will may dynamically change in time, attribution
of moral responsibility and guilt should always take into account
the entire history of previous choices made by the conscious subject.\\

Quantifying external coercion might be practically impossible for
the assessment of putative effects of complex environmental factors.
For example, a broker trading on the stock market may be influenced
by a variety of uncontrollable external stochastic factors to exhibit
so complicated behavior that it is unfeasible to deduce what could
have had happened in the absence of some controllable factor. The
measure given by \eqref{eq:coercion}, however, might be very useful
for quantifying the effects of certain drugs upon the performance
of humans engaged in some highly responsible professional activities that
require execution of free choices.
For example, any change from an initially unbiased probability of release ($p=0.5$) will decrease the amount of free will exercised and will exert some non-zero coercion per synapse. Volatile anesthetics severely decrease the probability of synaptic vesicle release ($p\to 0$) thereby erasing consciousness and free will \citep{Georgiev2018b}.
Alternatively, narcotic drugs such as cocaine may increase the probability of synaptic vesicle release ($p\to 1$) \citep{Venton2006,Buck2020} thereby also reducing free will.
Noteworthy, substances such as alcohol that induce failure of deterministic neurotransmission along the somatomotor pathways
could be viewed as infinitely coercive, and are therefore appropriately banned for consumption by drivers of motor vehicles.

\section{Discussion}
\label{sec:Discussion}

Philosophers have produced volumes of literature debating free will
in a binary fashion: a conscious agent either has free will or not.
Determinism of classical physics bans the existence of agents with
free will, which explains why the pursuit of compatibilism led to
redefining of ``free will'' as something else that has nothing to
do with the capacity of conscious agents to make genuine choices.
Resorting to redefinition of ``free will'', however, does not solve
the original problem and only helps create further confusion. Furthermore,
in a deterministic physical world there can be no meaningful discussion
of inherent biases or external coercion. Indeed, Hamilton's equations
\eqref{eq:Hamilton} imply that the dynamics of an agent is predetermined
with absolute certainty, hence always completely biased and without free will.
The latter fact precludes any attribution of responsibility
or guilt to physical agents and serves as a foundation for erecting
the domino theory of moral nonresponsibility. In classical physics,
human ethics becomes a mere historic accident, as it is impossible
to explain in what sense some moral value is objective. Modern quantum
physics, however, provides fundamentally indeterministic physical
laws that can naturally accommodate free will. When quantum physical systems
are measured, they have the capacity to choose a single measurement
outcome selected from a characteristic quantum probability distribution
obeying the Born rule. The average information gain from learning
the chosen measurement outcome, then serves as a quantitative indicator
of the amount of free will possessed by the quantum system.

The precise quantification of the amount of free will allowed by quantum
theory also illuminates old philosophical debates on the apparent
tension between our capacity to do otherwise and our desire to rationally
control our actions. If the physical laws were such that always either
maximal free will is granted or no free will is granted at all, then
we would not have had the opportunity to put ourselves in both of
these situations depending on the context. Quantum theory resolves
the tension by admitting varying amounts of free will for different measurement
contexts. The neurobiology of different free will contexts is as follows:
When the fundamental physical laws do not favor an outcome
of our actions, there is a quantum physical observable in the neural system whose measurement
outcomes are completely unbiased. For example, when we are born we
know nothing about the external world and there could be a synapse
whose probability of release of neurotransmitter is 50\% for Yes and
50\% for No. Without any inherent biases, we are maximally free to
choose to do one thing or the other. Suppose that we choose Yes and
the synapse releases neurotransmitter. If this decision leads to
benefits, the molecular machinery for the neural dopamine reward system
will be activated and the synapse may boost its probability for release
to say 75\%. Conversely, if the decision leads to harmful consequences,
the neural reward system will not be activated and the synapse may
decrease its probability for release to say 25\%. Thus, we can learn
from our previous successes or mistakes, trying to repeat the successes
but not to repeat the mistakes again. The effect of learning new information
upon an unbiased synapse will be to restrict the amount of manifested
free will. Indeed, we feel comfortable when we live in a risk free
environment. When everything goes smoothly in life, we do not want
to choose differently and do not need any free will. This may explain
the psychology behind denial of free will by accomplished philosophers.
When things become stressful in life, however, we urgently need our free will and the capacity to
choose otherwise. Subjected to high levels of adrenaline, synapses in the brain cortex that predictably activate or predictably remain silent, may be reset by experienced hardship
in life to the unbiased state 50\% for Yes and 50\% for No.
Thus, the presence or lack of relevant information could significantly impact
the amount of free will that we exercise with respect to open questions
for future action. The effect of learning upon free will could be
in either direction: to decrease the amount of free will when we perform
successfully or to enhance the amount of free will when we experience
hardship in life. The suggested effect of learning upon the amount
of synaptic free will is consistent with neurophysiological experiments,
which have already shown that neuronal electric activity or inactivity
is able to affect the probability of release of synaptic vesicles
at individual synapses \citep{Branco2008,Murthy2001}.

Classical philosophers such as Arthur Schopenhauer, who lived and
worked before the discovery of modern quantum physics in 1920s, viewed
determinism as an inescapable obstacle to genuine free will. Consequently,
they tried to redefine ``free will'' to indicate only willed or desired
actions, which may not have been free. This only confuses matters
because, although our emotions and desires influence our decisions,
sometimes we exercise our free will to choose things that we do not
desire. Also we do not have a persistent delusion that all bodily
actions are caused by us. For example, if someone strikes the patellar
ligament with a reflex hammer just below the knee cap, this will invariably
cause our leg to kick out, but the accompanying conscious experience
is as if the leg moved on its own without our conscious intention
to move it \citep{Georgiev2017}. Our subjective feeling of not using
our free will to move the leg is indeed consistent with neuroanatomy:
the patellar reflex is executed at the level of the spinal cord and
the electric signal arrives at the brain cortex only after the leg
motion has been already triggered. Thus, while the explanation of
the patellar reflex is wanting in the absence of free will, it is
quite simple if the existence of free will is acknowledged.

The presented quantum neurophysiology of free will vindicates the
trustworthiness of our introspective testimonies of which actions
we have freely chosen to do and which we have not. Because all newborns
lack any knowledge about the surrounding world, they exercise their
free will to explore various actions and experience their consequences.
Here is where the importance of the dopamine reward system and the
occurrence of pleasant or unpleasant feelings helps us learn through
trial-and-error mechanism. The free will that we are endowed with
allows us to choose what to become through our actions. We are born
ignorant into this world, but our lives are too short to let everyone
learn from his or her own mistakes. As a social species, we have contemplated
on the latter fact and reached to the conclusion that while the \emph{ignorance}
of an adult \emph{is not an excuse} for not bearing moral responsibility, we can excuse children
by transferring temporarily the responsibility from the child
to the parent who is engaged with the education of the child up to
a certain age. Admitting the existence of free will and contemplating
on social contexts in which the moral responsibility could be shared
by a more experienced member of the social group is the starting point
towards building an ethical system of moral values. In the animal
kingdom, there are a number of extraordinary examples in which grandparents
become responsible for the education of their grandoffspring, including
whales, dolphins, elephants, monkeys and humans \citep{Lee2003,Lee2016,Fairbanks1988a,Fairbanks1988b,Nakamichi2010,Lahdenpera2016,Coxworth2015,Nattrass2019,Hawkes1997,Kim2012,Coall2010}.
Of all animals, however, only humans teach their grandchildren abstract
ideas (such as free will) and leave as a legacy a set of moral values.
We hope that by raising the awareness of neuroscientists for the utility
of modern quantum physics, the free will problem will no longer be
intractable. The proposed quantitative measures of the amount of free
will $\mathcal{F}$ or the presence of external coercion $\mathcal{C}$
might be useful for the development of legal policies to control drug
use or for the attribution of guilt in legal cases.

\section*{Conflict of Interest}

The author certifies that he has no affiliations with or involvement in any organization or entity with any financial interest, or non-financial interest in the subject matter or materials discussed in this manuscript.

%\pagebreak
\appendix

\setcounter{secnumdepth}{0}
\section{Appendix: Problems in the philosophy of free will}

\label{app}

Quantum theory resolves many problems related to free will that were intractable in classical physics.

\paragraph{Locke's man in a locked room}
John Locke argued that our belief in free will could be an illusion sustained by our ignorance.
As an example, he considered a sleeping man who is transferred while asleep into
a locked room where a desirable companionship is presented. When the
man wakes up, he might choose to stay in the room by his own desire,
without being aware that the room is locked and there is no way out.
In this case, the man apparently stays in the room without having
the option to have done otherwise. Yet, because the man's own desire
has led him to choose the only available course of action, his ignorance
of the real situation will sustain the illusion of free will \citep{Locke1828}.
Although superficially plausible, Locke's argument is based on a conditional
statement and does not withstand further scrutiny. In fact, a real
person will sooner or later decide to exit the room, which means that
the illusion can persist only for a very limited period of time. In
real life, we never experience a desire to move one of our arms or
legs at a moment in time when the physical laws forbid such motion.
This could fit under the premisses of classical physics only if our desires are always aligned with
what is physically mandated by determinism thereby producing the illusion of free will.
However, it is easy to find real life situations in which we freely choose a course of action reluctantly
without any desire. For example, in a burning building a person may
have to choose between going through the fire sustaining severe skin
burns or jumping through high floor window sustaining severe bone
fractures. Neither of these two choices is in any sense desirable
and will never be experienced as desirable by the one who has to make them.
Thus, since free will allows us to choose undesirable courses
of action in real life, it is implausible that a physical mechanism can support the illusion of
free will in the absence of free will.

\paragraph{Frankfurt's advanced device}
Harry Frankfurt proposed a modern experiment
aimed at establishing compatibilism between free will and determinism.
He imagined an advanced device that is capable to continuously monitor
and, if needed, trigger electric activity in a subject's brain. With
such a device, it would be possible to do nothing if the subject chooses
to do some action~$A$, or trigger the device to electrically stimulate
the execution of action~$A$ if the subject has not chosen to do
the action~$A$. Similarly to Locke's argument, Frankfurt concludes
that the subject can choose to do the action~$A$, even though the
action~$A$ cannot be avoided \citep{Frankfurt1969}. The faulty reasoning
is again in the conditional character of the statement. If the experiment
is repeated several times, sooner or later the subject will chose
not to do the action~$A$ and be surprised that the action~$A$
is executed anyway. Thus, compatibilism cannot be true. Availability
of alternate possibilities is essential for the existence of free will.

\paragraph{Epiphenomenal belief in free will cannot evolve through natural selection}
Psychological experiments have found that cheating can be enhanced
in subjects who are given to read a written text passage stating that free
will is an illusion \citep{Vohs2008}. This has been interpreted as
an indication that the belief in free will decreases cheating, which
is a meaningful statement only under the implicit assumption that
our conscious beliefs can be causally effective in exerting an influence
on the processes that occur in the physical world. Those who believe
that free will is an illusion due to determinism have been fast to
adopt an evolutionary explanation for the origin of the free will
illusion, namely, we have evolved to believe in free will because
this make us nicer people in a social context. The latter statement
is demonstrably false, however, because conscious experiences are
epiphenomenal in deterministic functional theories of the human mind
\citep{Georgiev2013,Georgiev2017,Georgiev2020a,Georgiev2020b}. The
proof starts from the observation that conscious experiences do not
enter directly into Hamilton's equations \eqref{eq:Hamilton} of classical
physics. Therefore, conscious experiences can be generated by the
brain, but cannot change in any way the physical dynamics of the brain,
which is already fully determined by the classical physical quantities
that comprise the brain and enter directly into Hamilton's equations.
In other words, to predict the future dynamics of a deterministic
brain, we do not need to know whether it experiences something or
not. Instead, we just compute numerically what the Hamilton's equations
predict. Thus, the presence of conscious experiences in a deterministic
physical world cannot have any physical manifestation, which means
that consciousness can only be admitted as a causally ineffective
epiphenomenon. Epiphenomenal beliefs cannot evolve through natural
selection, which directly contradicts the claim that the belief in
free will makes us nicer human beings in a social context. In essence,
determinism forbids both free will and the evolutionary origin of
illusionary belief in free will.

\paragraph{Free will cannot evolve in a deterministic world}
Whether the physical world obeys deterministic or indeterministic physical laws is a universal
statement. Physical laws do not evolve in time, which implies that
free will is either allowed or forbidden at all times. Because natural
selection operates only on physically possible variations, it follows
that organisms cannot evolve free will in a deterministic physical
world that forbids free will. Nevertheless, some 21st century philosophers
have claimed that free will does evolve in a deterministic world.
Daniel Dennett has proposed a re-interpretation of free will in terms
of avoiding undesirable consequences \citep{Dennett2004}. For example,
a conscious agent could contemplate that bumping into a stone laying
on the road may break his leg and take avoiding action to circumvent
the stone. At first glance, such a definition agrees with the meaning of
free will, namely, avoiding an undesirable outcome that has non-zero
probability is a beneficial thing to do and certainly we will exercise our free will in the act of choosing the beneficial outcome.
In deterministic physical theories, however, the only allowable probabilities are zero or one. 
Therefore, given a classical model of the physical state of the world including ourselves, there are only two possible cases:
If the undesirable outcome occurs with probability of one,
then it is unavoidable and nothing that we can do will make any difference.
Similarly, if the undesirable outcome occurs with zero probability,
then it is impossible and again nothing that we can do will make any
difference. For example, consider the ancient Greeks who prayed to
Zeus, the god of the sky and thunder, not to strike them with a bolt
of lightning. Because Zeus does not exist, there is a zero probability
that Zeus strikes anybody with a lightning. Consequently, the prayers
of ancient Greeks cannot be beneficial to avoid something that is
impossible to occur anyway, and it would be meaningless to attribute
free will to those who prayed based on the fact that they have avoided
the impossible wrath of Zeus. Exactly the same will be the analysis
of avoiding the stone laying on the road in classical physics. Superficially,
the stone is a physical object and one could easily imagine how bumping
into a stone could break one's leg. In deterministic classical physics,
however, bumping into the stone given the exact physical state of
the world is either a solution of Hamilton's equations \eqref{eq:Hamilton},
hence unavoidable, or it is not a solution, hence impossible. If we
were to grant free will to physical processes that avoid the impossible,
then by definition all physical processes would possess free will
and the concept of free will would mean nothing. Noteworthy, our quantitative
measure of free will \eqref{eq:fw} is mathematically impervious to
superfluous addition of impossible outcomes due to the identity $0=0\log_{2}0$.

\paragraph{The randomness problem}
A common but misleading argument against
the relevance of quantum physics for the existence of free will is
to claim that if our choices were performed by flipping a fair coin,
then the resulting actions will be random manifestations of chance
or luck, hence they will be incompatible with free will. The mistake
in such argument is the conflation of \emph{external} and \emph{internal causes}
into a single category, namely, free will is defined by an inherent
propensity to perform choices, whereas the fair coin is an external
agent. Indeed, if a person flips a coin and executes the outcome chosen
by the coin, then the person manifests zero free will as the act of
copying the result from the coin is perfectly deterministic. The same
will be true if one films on a tape the behavior of a person endowed
with free will and then asks another person to copy the recorded behavior.
The second person manifests zero free will in the act of copying the
previously recorded behavior. A possible attempt to fix the argument
could be to insist that if the fair coin performs genuine unbiased
choices, then we should attribute free will to the fair coin itself.
But the answer is that this is exactly what we do, namely, if we attribute
free will to the brain by virtue of the physical laws, then we also
attribute free will to the constituent physical particles that build
up the brain. The free will does not pop in and out of existence in
violation of physical laws. In fact, in the evolutionary history of
the animal nervous systems, the narrative is reversed, that is the
brain possesses free will exactly because the physical components
from which it is built possess free will \citep{Georgiev2017}. The
latter position has been previously advocated by Conway and Kochen
who formalized their reasoning into a set of axioms from which they derived the so-called \emph{free will theorem}
\citep{Conway2006,Conway2009}.

\end{document}